\documentclass[iop,apj,tighten,twocolappendix,numberedappendix]{emulateapj}
\usepackage{apjfonts}

\usepackage{graphicx}
\usepackage{amsmath}
\usepackage{amssymb}
\usepackage{color}
\usepackage{mathtools}

\usepackage[breaklinks,colorlinks,citecolor=blue,linkcolor=red]{hyperref} 
\usepackage[all]{hypcap}

\newcommand\msun{\, \rm M_\odot}

\newcommand\zsun{\, \rm Z_\odot}
\newcommand\kms{\, \rm km\,s^{-1}}

\newcommand\vkick{{v_{\rm kick}}}
\newcommand\vesc{{v_{\rm esc}}}

\newcommand\be{\begin{equation}}
\newcommand\ee{\end{equation}}

\begin{document}

\title{On the Origin of GW190521-like events from repeated black hole mergers in star clusters}

\author{Giacomo Fragione\altaffilmark{1,2}, Abraham Loeb \altaffilmark{3}, Frederic A.\ Rasio\altaffilmark{1,2}}
 \affil{$^1$Center for Interdisciplinary Exploration \& Research in Astrophysics (CIERA), Evanston, IL 60202, USA} 
  \affil{$^2$Department of Physics \& Astronomy, Northwestern University, Evanston, IL 60202, USA}
  \affil{$^3$Astronomy Department, Harvard University, 60 Garden St., Cambridge, MA 02138, USA}

\begin{abstract}
LIGO and Virgo have reported the detection of GW190521, from the merger of a binary black hole (BBH) with a total mass around $150\msun$. While current stellar models limit the mass of any black hole (BH) remnant to about $40 - 50\msun$, more massive BHs can be produced dynamically through repeated mergers in the core of a dense star cluster. The process is limited by the recoil kick (due to anisotropic emission of gravitational radiation) imparted to merger remnants, which can escape the parent cluster, thereby terminating  growth. We study the role of the host cluster metallicity and escape speed in the buildup of massive BHs through repeated mergers. Almost independent of host metallicity, we find that a BBH of about $150\msun$ could be formed dynamically in any star cluster with escape speed $\gtrsim 200\kms$, as found in galactic nuclear star clusters as well as the most massive globular clusters and super star clusters. Using an inspiral-only waveform, we compute the detection probability for different primary masses ($\ge 60\msun$) as a function of secondary mass and find that the detection probability increases with secondary mass and decreases for larger primary mass and redshift. Future additional detections of massive BBH mergers will be of fundamental importance for understanding the growth of massive BHs through dynamics and the formation of intermediate-mass BHs.
\end{abstract}


\section{Introduction}
\label{sect:intro}

The detection of gravitational waves (GW) has revolutionized our understanding of black holes (BHs) and neutron stars (NSs). Since the first discovery, the LIGO and Virgo observatories have confirmed the detection of more than ten events \citep{aasi2015,acern2015,abb2019,abb2019b}. These observations have brought several surprises, including GW190412 \citep{ligo2020}, a binary black hole (BBH) merger with a mass ratio of nearly four-to-one, GW190814 \citep{ligo2020b}, a merger between a BH and a compact object of about $2.5\msun$, and GW190425 \citep{ligo2020c}, a merger of a binary NS of total mass nearly $3.4\msun$, the most massive binary NS observed so far.

The origin of binary mergers is still highly uncertain, with several possible scenarios that could potentially account for most of the observed events. These include mergers from isolated evolution of binary stars \citep{belcz2016,demi2016,giac2018}, dynamical assembly in dense star clusters \citep{askar17,baner18,fragk2018,rod18,sams18,ham2019,krem2019}, mergers in triple and quadruple systems induced through the Kozai-Lidov mechanism \citep{antoper12,ll18,fragg2019,flp2019,fragk2019,fragrasio2020}, and mergers of compact binaries in galactic nuclei \citep{bart17,sto17,rasskoc2019,mck2020}.

Another surprise is GW190521, a binary black hole (BBH) of total mass $\sim 150\msun$, consistent with the merger of two BHs with masses of $85^{+21}_{-14} \msun$ and $66^{+17}_{-18} \msun$ \citep{ligo2020new1,ligo2020new2}. Current stellar models predict a dearth of BHs both with masses larger than about $50\msun$ (high-mass gap) and smaller than about $5\msun$ (low-mass gap), with exact values depending on the details of the progenitor collapse \citep[e.g.,][]{fryer2012}. The high-mass gap results from the pulsational pair-instability process, which affects massive progenitors. Whenever the pre-explosion stellar core is in the range $45 - 65\msun$, large amounts of mass can be ejected, leaving a BH remnant with a maximum mass around $40 - 50\msun$ \citep{heger2003,woosley2017}. Therefore, GW190521 challenges our understanding of massive-star evolution.

BHs more massive than the limit imposed by pulsational pair-instability can be produced dynamically through repeated mergers of smaller BHs in the core of a dense star cluster, where three- and four-body interactions can catalyze the growth of a BH seed \citep[e.g.,][]{gultek2004}. A fundamental limit for repeated mergers comes from the recoil kick imparted to merger remnants as a result of anisotropic GW emission \citep{lou10,lou11}. Depending on the mass ratio and the spins of the merging objects, the recoil kick can be as high as $\sim 100 - 1000\kms$. If it exceeds the local escape speed, the merger remnant is ejected from the system and further growth is quenched. A number of studies have shown that massive globular clusters \citep[e.g.,][]{rodetal2019}, super star clusters \citep[e.g.,][]{rodr2020}, and nuclear clusters at the centers of galaxies \citep[e.g.,][]{anto2019,frsilk2020} are the only environments where the mergers of second- ($2$g) or higher-generation ($N$g) BHs could take place.

In this Letter, we explore the possibility that GW190521-like events (BBHs with total mass around $150\msun$) are the product of repeated mergers in a star cluster. The Letter is organized as follows. In Section~\ref{sect:limits}, we discuss the role of the cluster metallicity and escape speed in the assembly of massive BHs. In Section~\ref{sect:numer}, we discuss the assembly of massive BHs through repeated mergers in a variety of dynamically-active environments. In Section~\ref{sect:detect}, we discuss the detection probability for GW190521-like events. Finally, in Section~\ref{sect:concl}, we discuss the implications of our results and draw our conclusions.

\section{Limits on the hierarchical growth of Black Hole seeds}
\label{sect:limits}

Two main factors determine the ability of a BH seed to grow via repeated mergers: the environment metallicity and the host-cluster escape speed. The former sets the initial maximum seed mass, while the latter determines the maximum recoil kick that can be imparted to a merger remnant to be retained within the host cluster.

\subsection{Metallicity}

Dense star clusters form with a variety of initial masses, concentrations, and metallicities. Open clusters and super star clusters are high-metallicity environments \citep[e.g.,][]{portegies2010}, in contrast to most globular clusters \citep[e.g.,][]{harris1996}. Nuclear star clusters present both high- and low-metallicity stars, as a result of their complex history and various episodes of accretion and star formation \citep[e.g.,][]{anto2013}.

Metallicity is crucial in determining the maximum BH mass in a given environment. Low-metallicity systems can form BHs much more massive than high-metallicity systems. This difference is a result of stellar winds in massive stars. Higher-metallicity stars experience stronger winds and, as a consequence, larger mass-loss rates \citep{vink2001}, resulting in less massive BH progenitors prior to stellar collapse \citep{spera2017}. Therefore, typical globular clusters are expected to produce more massive BHs than open and super star clusters. 

To demonstrate the role of metallicity, we consider a sample of stars in the mass range of BH progenitors, $[20\msun-150\msun]$, and evolve them using the stellar evolution code \texttt{SSE} \citep{hurley2000,hurley2002}. We use the updated version of \texttt{SSE} from \citet{banerjee2020}, with the most up-to-date prescriptions for stellar winds and remnant formation; it produces remnant populations consistent with those from \texttt{StarTrack} \citep{belc2008}. We choose four different values of the metallicity $Z$, namely $0.01\zsun$, $0.1\zsun$, $0.5\zsun$, and $\zsun$.

\begin{figure} 
\centering
\includegraphics[scale=0.6]{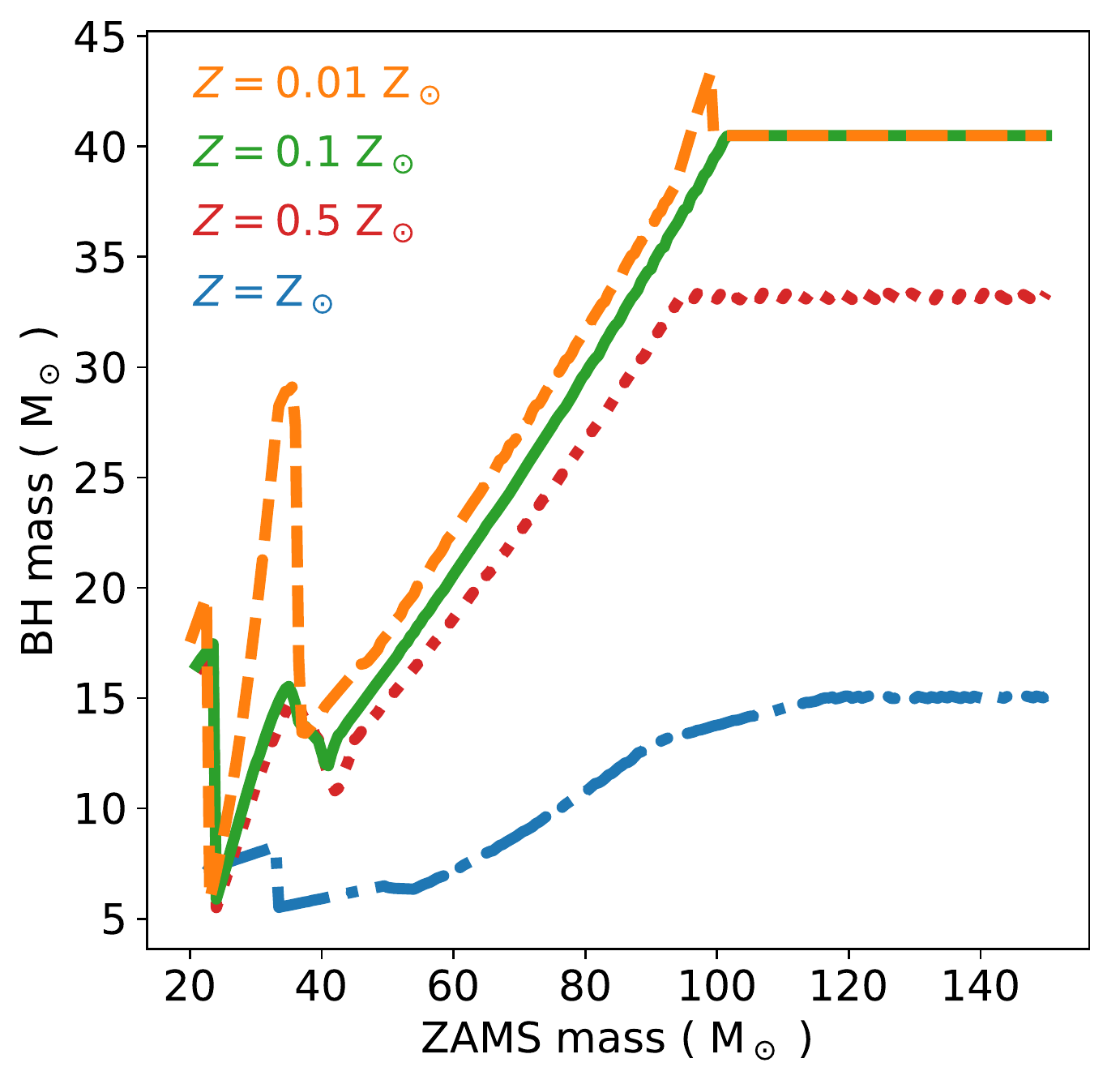}
\caption{BH mass as a function of zero-age main-sequence (ZAMS) mass for single stars, computed using \texttt{SSE} \citep{hurley2000,hurley2002} with updates from \citet{banerjee2020}. The four colors denote the four different metallicities.}
\label{fig:sse}
\end{figure}

In Figure~\ref{fig:sse}, we show the final BH mass as a function of the zero-age main sequence (ZAMS) mass for single stars computed using \texttt{SSE}, for different metallicities. For solar metallicity, the maximum BH mass is of about $15\msun$, and this increases to about $35\msun$ and $45\msun$ for $Z=0.5\zsun$ and $Z=0.1\zsun$, respectively. Note that, for the stellar-mass range considered, metallicities lower than about $0.1\zsun$ would all produce very similar initial BH mass functions.

Metallicity, therefore, limits the initial mass of the BH seed that can undergo repeated mergers. At the same time, it constrains also the maximum mass of the BHs the seed can merge with. In a low-metallicity cluster, GW190521 components could be 2g BHs, each the remnant of a merger of 1g BHs. On the other hand, they could be 4g or 5g BHs if their progenitors were born in a solar-metallicity environment. 

We note in passing that runaway growth of a very massive star could also be triggered through physical collisions if the initial density of the host cluster is sufficiently high. Such a star could eventually collapse to form a BH in the high-mass gap, or even an intermediate-mass BH \citep{porte2004,gurk2006,pan2012,krgap2020}. However, stellar evolution is highly unconstrained in this regime. Some models suggest that only a low-metallicity star with mass $\gtrsim 200\msun$ could directly collapse to a BH more massive than about $80\msun$ \citep{spera2017,renzo2020}.

\subsection{Recoil kicks}

The escape speed $v_{\rm esc}$ from the core of a star cluster is determined by its mass and density profile. The more massive and dense the cluster is, the higher the escape speed. Open clusters, globular cluster, and nuclear star clusters have typical escape speeds $\sim 1\kms$, $\sim 10\kms$, and $\sim 100\kms$, respectively. Note, however, that the escape speed of a given environment may change over time depending on the details of its formation history and dynamical evolution \citep[see e.g. Fig.~3][]{rodr2020}.

\begin{figure*} 
\centering
\includegraphics[scale=0.55]{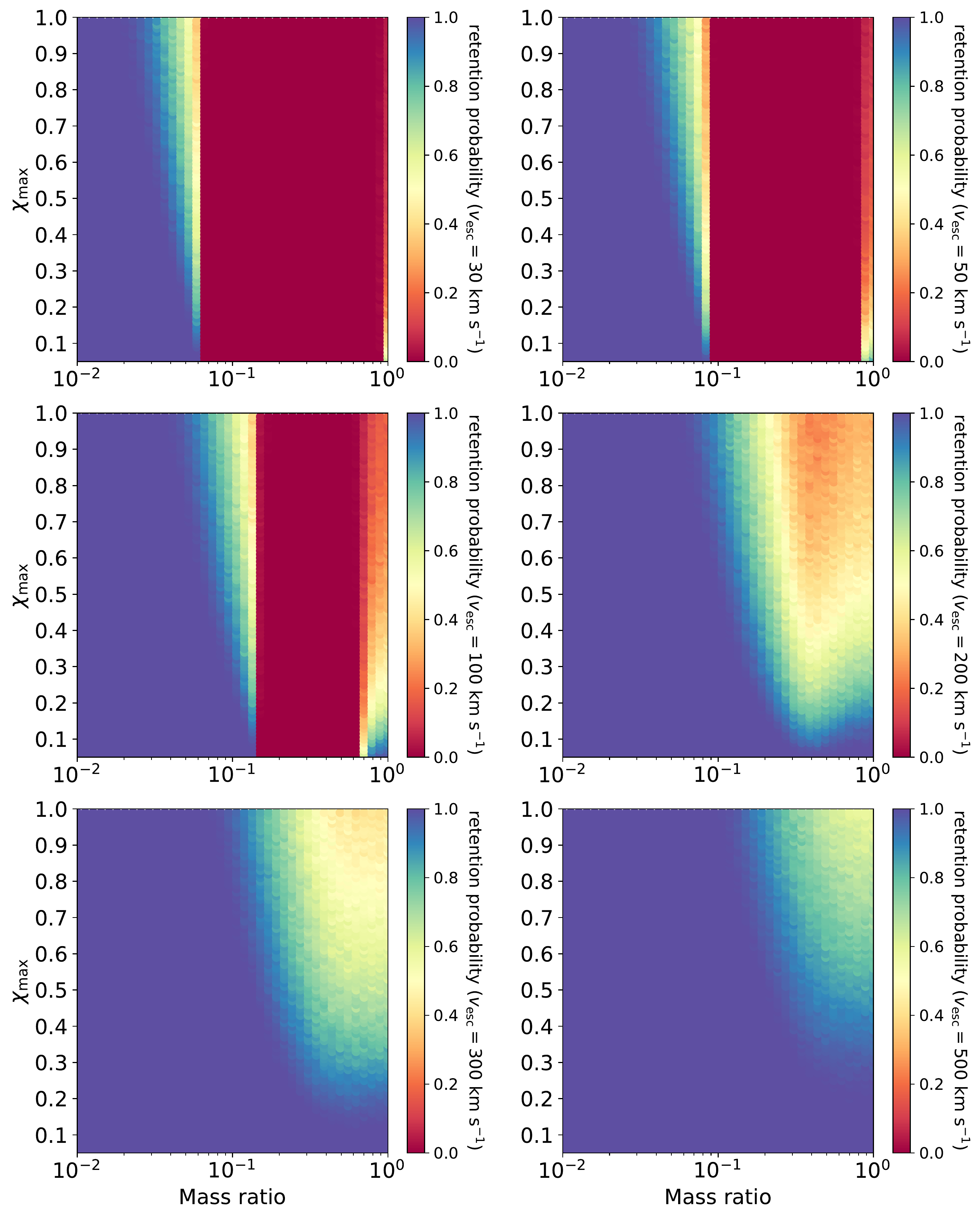}
\caption{Probability to retain the merger remnant of a BBH as a function of the BBH mass ratio and BH spins, for different cluster escape speeds: $30\kms$ (top-left), $50\kms$ (top-right), $100\kms$ (center-left), $200\kms$ (center-right), $300\kms$ (bottom-left), $500\kms$ (bottom-right). BH spins are drawn from a uniform distribution in the range $[0,\chi_{\max}]$.}
\label{fig:probret}
\end{figure*}

Due to the anisotropic emission of GWs at merger, a recoil kick is imparted to the merger remnant \citep{lou12}. The host escape speed then determines the fraction of retained remnants. The recoil kick depends on the asymmetric mass ratio $\eta=q/(1+q)^2$, where $q=m_2/m_1<1$ ($m_1$ and $m_2$ are the masses of the merging BHs), and on the magnitude of the dimensionless spins, $|\mathbf{{\chi_1}}|$ and $|\mathbf{{\chi_2}}|$ (corresponding to $m_1$ and $m_2$). We model the recoil kick following \citep{lou10} as
\begin{equation}
\textbf{v}_{\mathrm{kick}}=v_m \hat{e}_{\perp,1}+v_{\perp}(\cos \xi \hat{e}_{\perp,1}+\sin \xi \hat{e}_{\perp,2})+v_{\parallel} \hat{e}_{\parallel}\,,
\label{eqn:vkick}
\end{equation}
where
\begin{eqnarray}
v_m&=&A\eta^2\sqrt{1-4\eta}(1+B\eta)\\
v_{\perp}&=&\frac{H\eta^2}{1+q}(\chi_{2,\parallel}-q\chi_{1,\parallel})\\
v_{\parallel}&=&\frac{16\eta^2}{1+q}[V_{1,1}+V_A \tilde{S}_{\parallel}+V_B \tilde{S}^2_{\parallel}+V_C \tilde{S}_{\parallel}^3]\times \nonumber\\
&\times & |\mathbf{\chi}_{2,\perp}-q\mathbf{\chi}_{1,\perp}| \cos(\phi_{\Delta}-\phi_{1})\,.
\end{eqnarray}
The $\perp$ and $\parallel$ refer to the directions perpendicular and parallel to the orbital angular momentum, respectively, while $\hat{e}_{\perp,1}$ and $\hat{e}_{\perp,2}$ are orthogonal unit vectors in the orbital plane. We have also defined the vector
\begin{equation}
\tilde{\mathbf{S}}=2\frac{\mathbf{\chi}_{2,\perp}+q^2\mathbf{\chi}_{1,\perp}}{(1+q)^2}\,,
\end{equation}
$\phi_{1}$ as the phase angle of the binary, and $\phi_{\Delta}$ as the angle between the in-plane component of the vector
\begin{equation}
\mathbf{\Delta}=M^2\frac{\mathbf{\chi}_{2}-q\mathbf{\chi}_{1}}{1+q}
\end{equation}
and the infall direction at merger. Finally, we adopt $A=1.2\times 10^4$ km s$^{-1}$, $H=6.9\times 10^3$ km s$^{-1}$, $B=-0.93$, $\xi=145^{\circ}$ \citep{gon07,lou08}, and $V_{1,1}=3678$ km s$^{-1}$, $V_A=2481$ km s$^{-1}$, $V_B=1793$ km s$^{-1}$, $V_C=1507$ km s$^{-1}$ \citep{lou12}. The final total spin of the merger product and its mass are computed following \citet{rezzolla2008}. 

In Figure~\ref{fig:probret}, we show the probability (over $10^4$ realizations) to retain the merger remnant of a BBH as a function of the BBH mass ratio ($q$) for different cluster escape speeds: $30\kms$ (top-left), $50\kms$ (top-right), $100\kms$ (center-left), $200\kms$ (center-right), $300\kms$ (bottom-left), $500\kms$ (bottom-right). We sample BH spins from a uniform distribution in the range $[0,\chi_{\max}]$. The recoil kick depends crucially on the maximum intrinsic spin of the merging BHs; while for low spins $v_{\rm kick}\sim 100\kms$, for high spins $v_{\rm kick}\sim 1000\kms$ \citep[e.g.,][]{holl2008,fgk2018,gerosa2019,anto2019,fragrasio2020,mapell2020}. The mass ratio also plays an important role, with the recoil kick decreasing significantly in magnitude for $q\lesssim 0.1$ (both for spinning and non-spinning BHs) and for $q\gtrsim 0.9$ (non-spinning BHs). Clusters with low-escape speeds ($v_{\rm esc}\lesssim 100\kms$) can only retain the merger products of very unequal-mass binaries ($q\lesssim 0.1$) and the remnants of roughly equal-mass BBH mergers with low-spinning components. On the other hand, clusters with larger escape speeds ($v_{\rm esc}\gtrsim 100\kms$) can retain, with various probabilities, remnants of various mass-ratio and spins. The remnants of the merger of highly-spinning equal-mass BBHs could even be ejected in very massive and dense clusters ($v_{\rm esc}\approx 500\kms$).

The host cluster escape speed plays a crucial role in the growth of a BH seed through repeated mergers. Typical small open clusters ($v_{\rm esc}\sim 1\kms$) do not provide the right environment for growth of a BH seed. If BHs are born with low spins \citep{fullerma2019}, the recoil kick could be small enough to retain a merger product within a typical globular cluster ($v_{\rm esc}\sim 10\kms$). However, if BHs are born with high spins, only more massive and denser systems, such as nuclear star clusters ($v_{\rm esc}\sim 100\kms$), could retain the remnant. In a low-metallicity cluster, GW190521 components could be 2g BHs, remnants of the mergers of nearly equal-mass 1g BHs. To retain them in a cluster with $v_{\rm esc}\lesssim 200\kms$, the progenitors should have been born with low spins. Otherwise, the two components of GW190521 could have been formed through repeated mergers of a massive 1g BH ($\gtrsim 40\msun$) with low-mass 1g BHs ($\lesssim 10\msun$) in a nuclear star cluster ($v_{\rm esc}\gtrsim 200\kms$). On the other hand, in a high-metallicity environment, GW190521 components could be 4g or 5g BHs, since the maximum 1g BH mass is limited to about $15\msun$. Therefore, they should be retained after several mergers. Since there is a negligible probability to retain a BH remnant in the region $0.2\lesssim q\lesssim 0.8$ for $v_{\rm esc}\lesssim 200\kms$, only nuclear star clusters or the most massive globular clusters and super star clusters ($v_{\rm esc}\gtrsim 200\kms$) could still form GW190521.

\section{GW190521-like events from repeated mergers}
\label{sect:numer}

\begin{figure*} 
\centering
\includegraphics[scale=0.55]{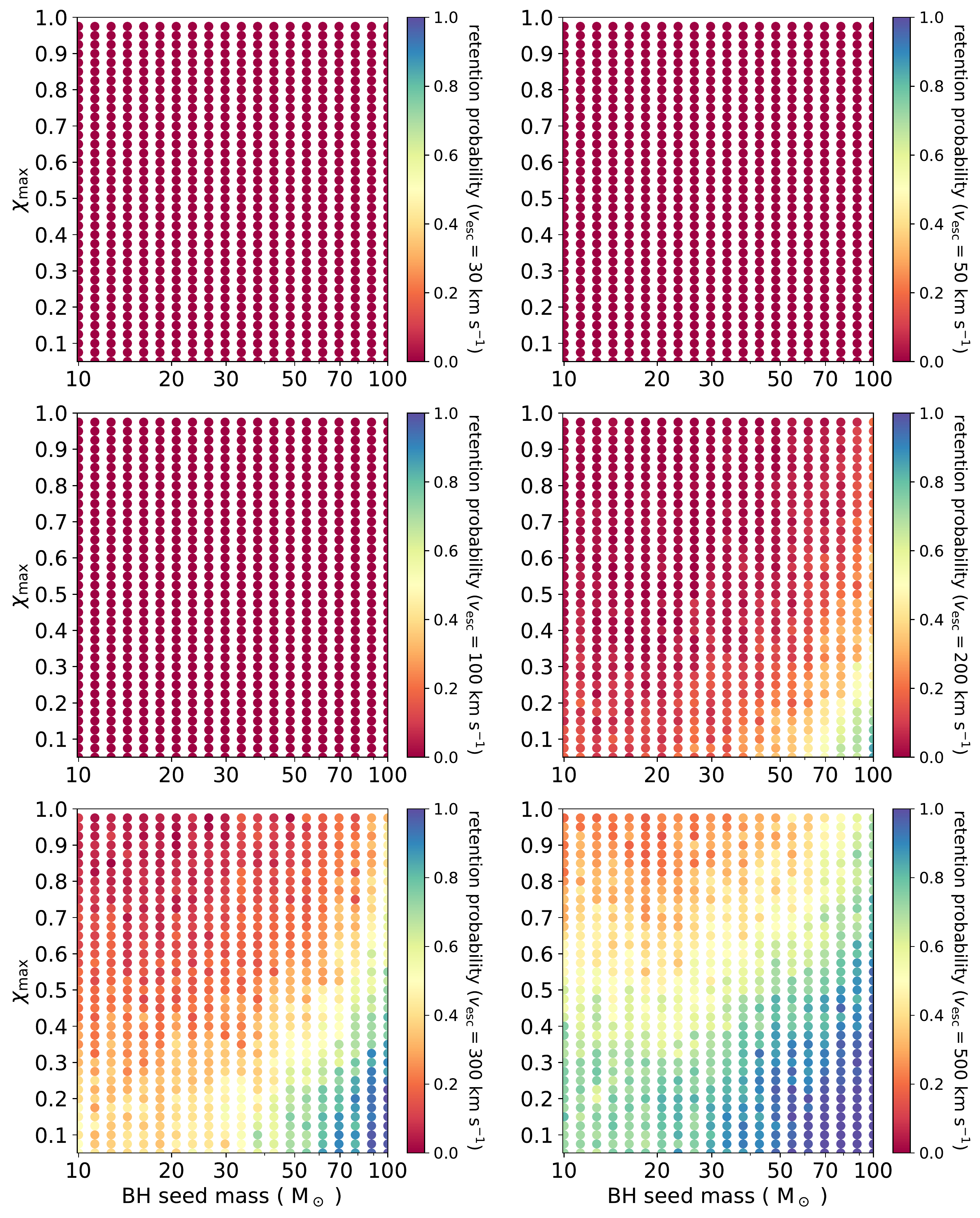}
\caption{Probability of forming a BBH of total mass $150\msun$ through successive mergers as a function of the BH seed mass for different cluster escape speeds: $30\kms$ (top-left), $50\kms$ (top-right), $100\kms$ (center-left), $200\kms$ (center-right), $300\kms$ (bottom-left), $500\kms$ (bottom-right). BH spins are drawn from a uniform distribution in the range $[0,\chi_{\max}]$. The cluster metallicity is fixed to $Z=0.01\zsun$.}
\label{fig:probret2a}
\end{figure*}

\begin{figure*} 
\centering
\includegraphics[scale=0.55]{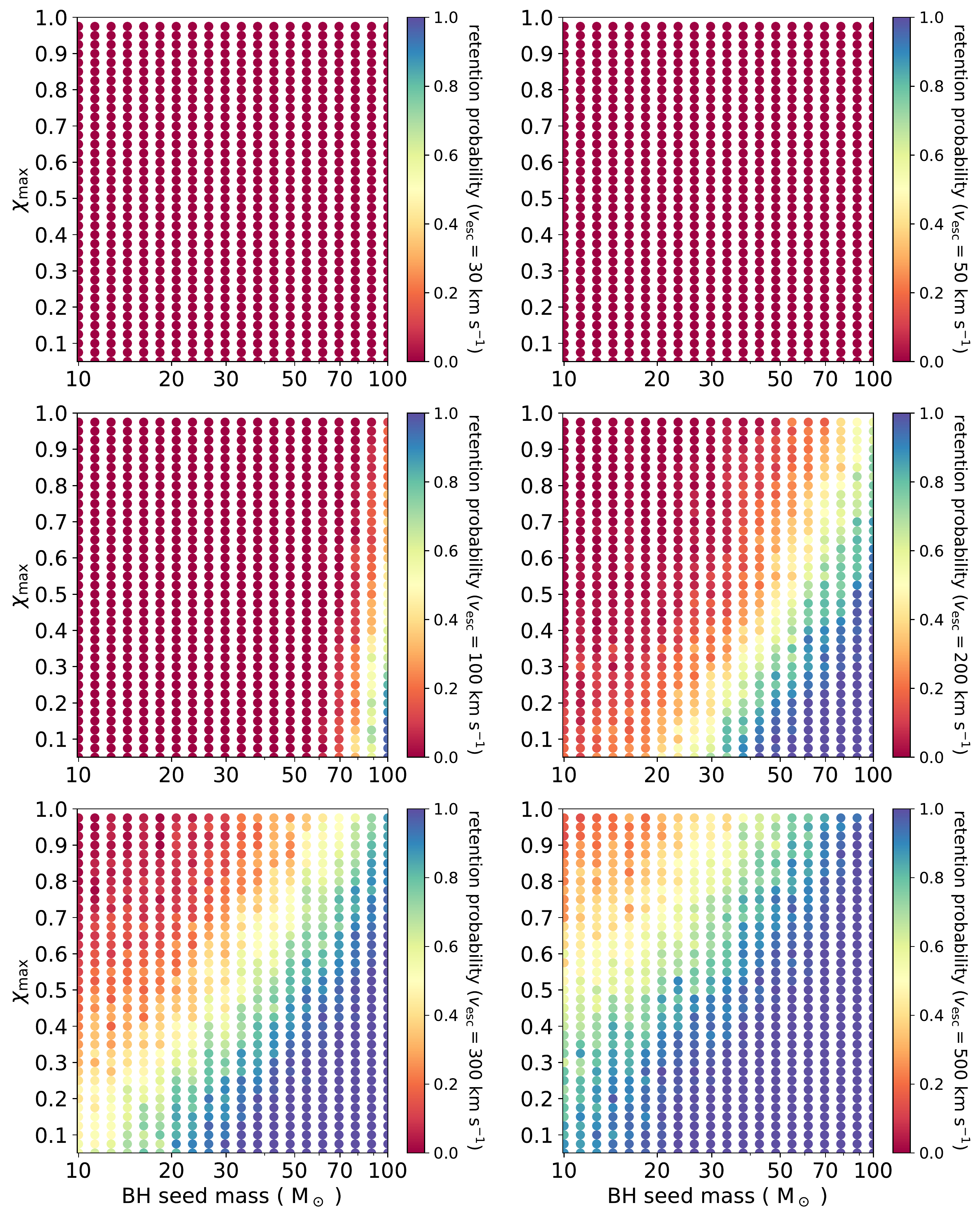}
\caption{Same as Figure~\ref{fig:probret2a}, but for $Z=\zsun$.}
\label{fig:probret2b}
\end{figure*}

GW190521 is a remarkable event since both of its components are likely the remnant of a previous BBH merger \citep[see also][]{ligo2020new2}. In this Section, we discuss the formation of GW190521-like events, requiring that a binary of total mass $150\msun$ be formed through repeated mergers of a growing BH seed within a cluster of escape speed $v_{\rm esc}$.

We run $10^4$ Monte Carlo experiments, where we simulate the growth of a BH seed via repeated mergers. After each merger, we compute the recoil kick using Eq.~\ref{eqn:vkick}. If $\vkick>v_{\rm esc}$, we consider the BH ejected from the system and further growth is impossible; otherwise, we proceed with generating a new merger event. In our numerical experiment, the probability of forming a BBH of $150\msun$ depends mainly on four parameters:
\begin{enumerate}
    \item the cluster metallicity $Z$, which fixes the maximum initial seed BH mass and the maximum mass of the BHs it can merge with;
    \item the steepness of the pairing probability for BHs in binaries that merge, $\propto (m_1+m_2)^\beta$, which sets the secondary mass;
    \item the maximum spin $\chi_{\max}$, which affects the maximum recoil kick;
    \item the escape speed from the host cluster $v_{\rm esc}$, which fixes the maximum kick velocity for the remnant to be retained within the host cluster.
\end{enumerate}
In our study, we choose two values of the metallicity, $Z=\zsun$ and $Z=0.01\zsun$, which fix the maximum seed mass to about $15\msun$ and $45\msun$, respectively. Note that, for the stellar-mass range considered, metallicities lower than about $0.1\zsun$ would all produce very similar initial BH mass functions \citep{belcz2010}. However, as mentioned above, collisions and mergers of massive stars could produce a BH remnant in the high-mass gap, or even an intermediate-mass BH \citep{porte2004,gurk2006,krgap2020}. To explore this possibility, we also consider BH seed masses up to $100\msun$. In our models, the cluster metallicity only sets the maximum mass for the BHs the seed can merge with. We sample the intrinsic spins at birth of BHs from a uniform distribution in the range $[0,\chi_{\max}]$, with $0\le \chi_{\max} \le 1$. We set $\beta=4$, as appropriate for binaries formed via dynamical three-body processes \citep{olear2016}. Finally we consider $\vesc$ in the range $[30\kms,500\kms]$ to encompass the full range of star clusters, from small open clusters to very massive nuclear star clusters.

Figure~\ref{fig:probret2a} shows the probability to form a BBH of total mass $150\msun$ as a function of the seed mass for different cluster escape speeds. We set the cluster metallicity to $Z=0.01\zsun$. A crucial role is played by $\chi_{\max}$. The probability of forming a BBH of total mass of about $150\msun$ is nearly $3$--$4$ times larger when $\chi_{\max}=0.2$ than when $\chi_{\max}=1$, even for clusters with large escape speeds. We find that clusters with escape speeds $\lesssim 50\kms$ cannot assemble such a massive BBH since the recoil kick is too large to retain a growing BH seed, independent of the maximum spin at birth. Clusters with escape speeds of $100\kms$ can form a massive BBH with total mass $150\msun$ only for large initial seed masses, $\gtrsim 70\msun$, and low-spins, with $\chi_{\max} < 0.4$. Only star clusters with $v_{\rm esc}\gtrsim 200\kms$ could form a BBHs of $150\msun$ starting from a highly-spinning BH seed of mass $\lesssim 50\msun$, which is consistent with current stellar evolutionary models for $Z=0.01\zsun$.

In Figure~\ref{fig:probret2b}, we explore the same parameter space but with $Z=\zsun$. As a general trend, solar metallicity favors the formation of a massive BBH for a wider portion of the parameter space, since the maximum BH mass is limited to $15\msun$, thus producing mostly mergers with low mass ratios. This, in turn, leads to lower recoil kicks imparted to the merger remnant, which can be retained more easily. However, if the BH seed mass is limited to $15\msun$, about the maximum mass allowed by stellar evolutionary models at solar metallicity, the formation of a BBH of about $150\msun$ is probable only for $v_{\rm esc}\gtrsim 200\kms$.

We have also run models where we consider $\beta=3$ and $\beta=5$, to study the role of the steepness of the pairing probability for BBHs that merge. We find no significant difference from the case $\beta=4$.

\section{Detection probability for GW190521-like events}
\label{sect:detect}

\begin{figure*} 
\centering
\includegraphics[scale=0.55]{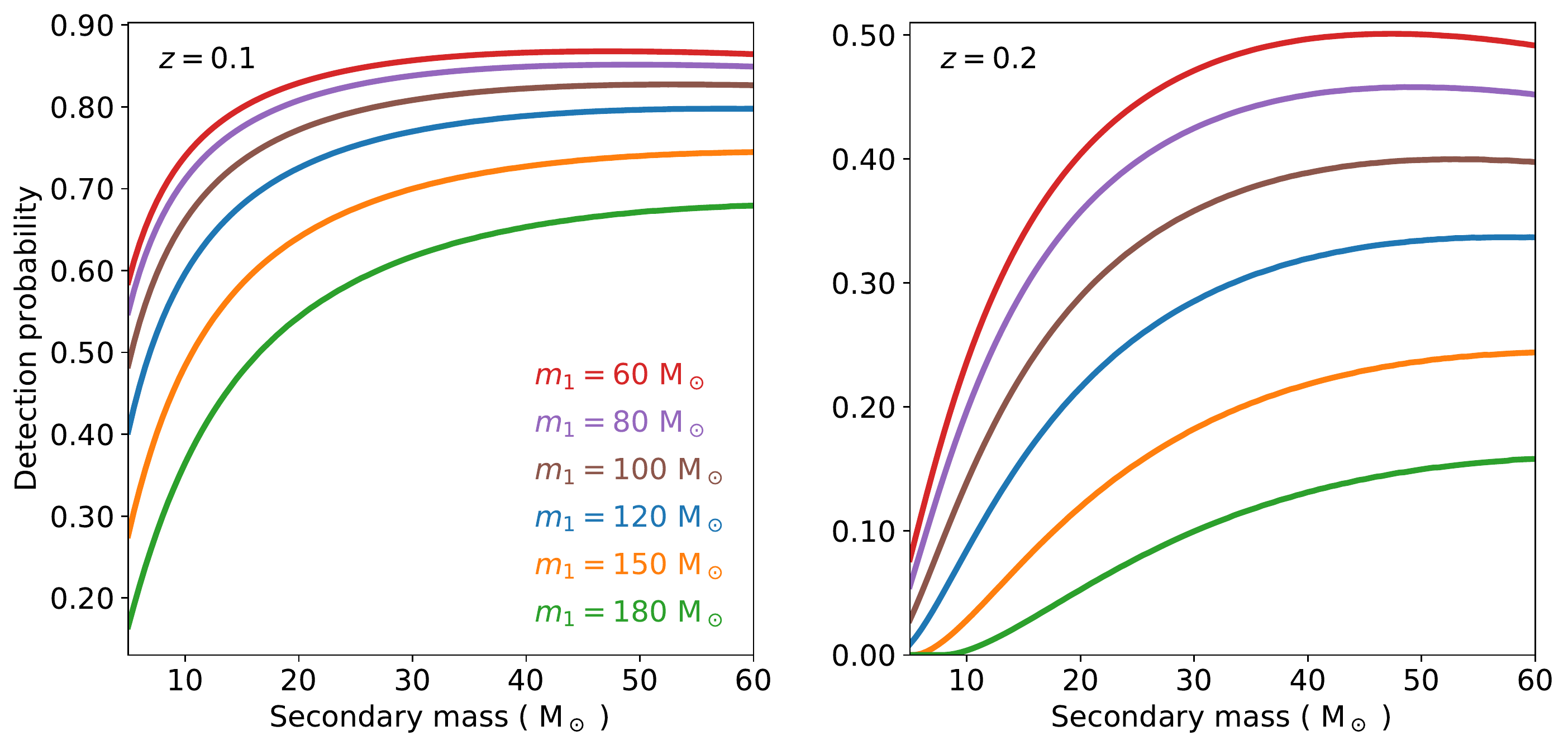}
\caption{Detection probability (Eq.~\ref{eqn:detec}) for different primary masses ($\ge 60\msun$) as a function of the secondary mass, assuming a signal-to-noise ratio threshold $\rho_{\rm thr}=8$ and a single LIGO instrument at design sensitivity. Different colors represent different primary masses. Left panel: source redshift $z=0.1$; right panel: source redshift $z=0.2$.}
\label{fig:pdetec}
\end{figure*}

We now consider the probability of detecting a BBH merger where one of the components ($m_1$) is a massive BH.

For a source with masses $m_1$ and $m_2$, merging at a luminosity distance $D_{\rm L}$, the signal-to-noise ratio (S/N) can be expressed relative to the strain noise spectrum of a single interferometer $S_{\rm n}(f)$ and the Fourier transform $\tilde{h}(f)$ of the GW strain received at the detector by an arbitrarily oriented and located source as \citep{oshau2010}
\begin{equation}
\rho=\sqrt{4 w^2\int_{0}^{f_{\rm ISCO}} \frac{|\tilde{h}(f)|^2}{S_n(f)} df}\,,
\label{eqn:rhof}
\end{equation}
where $w$ is a purely geometrical (and S/N-threshold-independent) function \citep[see Eq.~2 in][]{oshau2010}, which takes values between $0$ and $1$, and completely encompasses the detector- and source-orientation-dependent sensitivity, $f_{\rm ISCO}=c^3/(6^{1.5}\pi GM)$ is the ISCO frequency, and $|\tilde{h}(f)|$ is the frequency-domain waveform amplitude \citep[e.g., Eq.~3 in][]{abadie2010}
\begin{equation}
|\tilde{h}(f)|=\sqrt{\frac{5}{24\pi^{4/3}}} \frac{G^{5/6}}{c^{3/2}} \frac{M_{\rm c,z}^{5/6}}{D_{\rm L}(1+z)f_{\rm GW,z}^{7/6}} \,.
\label{eqn:htilde}
\end{equation}
In the previous equation, $f_{\rm GW,z}$ is the observed (detector frame) frequency, related to the binary orbital frequency by $f_{\rm GW,z}(1+z)=f_{\rm orb}$, $M_{\rm c,z}$ is the redshifted chirp mass, related to the rest-frame chirp mass by $M_\mathrm{c}=M_\mathrm{c,z}(1+z)$, and
\begin{equation}
D_{\rm L}=(1+z)\frac{c}{H_0}\int_{0}^z \frac{d\zeta}{\sqrt{\Omega_{\rm M}(1+\zeta^3)+\Omega_\Lambda}}\,,
\end{equation}
where $z$ is the redshift and $c$ and $H_0$ the velocity of light and Hubble constant \footnote{We set $\Omega_{\rm M}=0.286$ and $\Omega_\Lambda=0.714$ \citep{planck2016}.}, respectively. For LIGO/Virgo we adopt a noise model from the analytical approximation of Eq.~4.7 in \citet{aji2011}
\begin{eqnarray}
S_n(f)&=&10^{-48}\,{\rm Hz}^{-1}(0.0152 x^{-4}+0.2935x^{9/4}+2.7951x^{3/2})\nonumber\\
&-&6.5080x^{3/4}+17.7622)\,,
\end{eqnarray}
where $x=f/245.4\ {\rm Hz}$, which is in excellent agreement with the publicly available Advanced LIGO design noise curve \footnote{\url{https://dcc.ligo.org/LIGO-T0900288/public}}.

The detection probability $p_{\rm det}(m_1,m_2,z)$ is simply the fraction of sources of a given mass located at the given redshift that exceeds the detectability threshold in S/N, assuming that sources are uniformly distributed in sky location and orbital orientation, defined as \citep[e.g.][]{domin2015}
\begin{equation}
p_{\rm det}(m_1,m_2,z)=P(\rho_{\rm thr}/\rho_{\rm opt})\,,
\label{eqn:detec}
\end{equation}
where $\rho_{\rm opt}=\rho(w=1)$. A good approximation is given by Eq.~12 in \citet{domin2015}
\begin{eqnarray}
P(\mathcal{W})&=&a_2(1-\mathcal{W}/\alpha)^2+a_4(1-\mathcal{W}/\alpha)^4\nonumber\\
&+&a_8(1-\mathcal{W}/\alpha)^8+(1-a_2-a_4-a_8)(1-\mathcal{W}/\alpha)^{10}\,,
\end{eqnarray}
where $a_2=0.374222$, $a_4=2.04216$, $a_8=-2.63948$, and $\alpha=1.0$. We assume $\rho_{\rm thr}=8$.

In Figure~\ref{fig:pdetec}, we show the detection probability (Eq.~\ref{eqn:detec}) for different primary masses ($m_1\ge 60\msun$) as a function of the secondary mass, assuming a signal-to-noise-ratio threshold $\rho_{\rm thr}=8$ \citep{ligo2016} and a single LIGO instrument at design sensitivity \citep{ligo2018}. We represent different primary masses in different colors, and fix the source redshift at $z=0.1$ (left panel) and $z=0.2$ (right panel). We find that the larger the secondary mass the larger the detection probability, while it decreases for larger primary masses. For $m_1=60\msun$, we find that the the detection probability is in the range $60\%$--$90\%$ at $z=0.1$, which decreases to about $20\%$--$60\%$ for $m_1=180\msun$. As expected, a larger redshift also leads to smaller detection probabilities, which are a factor of about $2$--$6$ smaller in the case $z=0.2$.

Note that we have used an inspiral-only waveform and have not taken into account the S/N from merger and ringdown, which could important for high-mass binaries, as GW190521 \citep[see e.g.,][]{khan2016}.

\section{Discussion and conclusions}
\label{sect:concl}

GW190521 challenges our current understanding of stellar evolution for massive stars. Stellar models predict that whenever the pre-collapse stellar core is approximately in the range $45\msun$--$65\msun$, large amounts of mass can be ejected following the onset of the pulsational pair-instability process, leaving a BH remnant with a maximum mass around $40\msun$--$50\msun$ \citep{woosley2017}. Since only a rare star of extremely low-metallicity and mass $\gtrsim 200\msun$ could collapse to a BH of mass $\gtrsim 80\msun$ \citep{spera2017,renzo2020}, GW190521 is unlikely to have been born as an isolated binary.

BHs more massive than the limit imposed by pulsational pair-instability could be produced dynamically through repeated mergers of smaller BHs in the core of a dense star cluster. However, the recoil kick imparted to the merger remnant, which crucially depends on the BBH mass ratio and the distribution of BH spins at birth, could eject it out of the parent cluster, terminating growth \citep{anto2019,frsilk2020}.

We have simulated the growth of massive BHs starting from different BH seeds, as a function of the maximum BH spin, $\chi_{\max}$, in host star clusters with various metallicities and escape speeds. We have found that the probability of forming GW190521-like events with total mass around $150\msun$ depends crucially on the maximum BH spin at birth. The probability of forming such massive BBHs is about $3$ times larger with $\chi_{\max}=0.2$ than with $\chi_{\max}=1$, even for clusters with large escape speeds. Almost independent of metallicity, we have demonstrated that only nuclear star clusters or the most massive globular clusters and super star clusters could form BBHs with total mass around $150\msun$. This conclusion does not change when higher-mass seeds ($\gtrsim 50\msun$) are considered.

If GW190521 was formed in a low-metallicity cluster, such as an old globular cluster, its components could be 2g BHs, remnants of previous mergers of nearly equal-mass 1g BHs. We have shown that in a cluster with $v_{\rm esc}\lesssim 200\kms$, the progenitor 1g BHs must then have been born with low spins. Otherwise, the two components of GW190521 could have been formed through repeated minor mergers of a massive 1g BH with low-mass 1g BHs ($\lesssim 10\msun$) in a nuclear star cluster ($v_{\rm esc}\gtrsim 200\kms$). On the other hand, if GW190521 was born in a high-metallicity environment, its components could be 4g or 5g BHs, which have to be retained after several mergers. Since there is a negligible probability of retaining a remnant for $0.2\lesssim q\lesssim 0.8$ for $v_{\rm esc}\lesssim 200\kms$, we have demonstrated that only a nuclear star cluster or the most massive globular clusters and super star clusters (with $v_{\rm esc}\gtrsim 200\kms$) could form GW190521.

We have also computed the detection probability for different primary masses ($\ge 60\msun$) as a function of the secondary mass, assuming a signal-to-noise-ratio threshold $\rho_{\rm thr}=8$ \citep{ligo2016} and a single LIGO instrument at design sensitivity \citep{ligo2018}. We have found that the larger the secondary mass is, the larger the detection probability becomes. On the other hand the detection probability decreases for larger primary masses and redshifts.

GW190521 is a remarkable event that challenges our current theoretical understanding of BBH formation, opening debates about its origin and detection \citep[e.g.,][]{deluca2020,fish2020,gaya2020,liubro2020,liulai2020,rice2020,romero2020,safa2020,sakst2020,sams18}. Future detections of such massive mergers will help constrain our models for the growth of massive BHs through stellar dynamics and the formation of intermediate-mass BHs \citep{fraga2018,fragb2018,frbr2019,greene2019}.

\section*{Acknowledgements}

GF acknowledges support from a CIERA Fellowship at Northwestern University. FAR acknowledges support from NSF Grant AST-1716762. This work was supported in part by Harvard's Black Hole Initiative, which is funded by grants from JFT and GBMF.

\bibliographystyle{yahapj}
\bibliography{refs}

\begin{thebibliography}{}
\providecommand\natexlab[1]{#1}
\providecommand\JournalTitle[1]{#1}

\bibitem[{{Aasi} {et~al.}(2015){Aasi}, {Abbott}, {Abbott}, {Abbott},
  {Abernathy}, {Ackley}, {Adams}, {Adams}, {Addesso}, {Adhikari}, {Adya},
  {Affeldt}, {Aggarwal}, {Aguiar}, {Ain}, {Ajith}, {Alemic}, {Allen},
  {Amariutei}, {Anderson}, {Anderson}, {Arai}, {Araya}, {Arceneaux}, {Areeda},
  {Ashton}, \& {Ast}}]{aasi2015}
{Aasi}, J., {Abbott}, B.~P., {Abbott}, R., {et~al.} 2015,
  \href{http://dx.doi.org/10.1088/0264-9381/32/7/074001}{\JournalTitle{Classical
  and Quantum Gravity}, 32, 074001}

\bibitem[{{Abadie} {et~al.}(2010)}]{abadie2010}
{Abadie}, J., {et~al.} 2010,
  \href{http://dx.doi.org/10.1088/0264-9381/27/17/173001}{\JournalTitle{Classical
  and Quantum Gravity}, 27, 173001}

\bibitem[{{Abbott} {et~al.}(2019{\natexlab{a}}){Abbott}, {Abbott}, {Abbott},
  {Abraham}, {Acernese}, {Ackley}, {Adams}, {Adhikari}, {Adya}, {Affeldt},
  {Agathos}, {Agatsuma}, {Aggarwal}, {Aguiar}, {Aiello}, {Ain}, {Ajith},
  {Allen}, {Allocca}, {Aloy}, {Altin}, {Amato}, {Ananyeva}, {Anderson}, {LIGO
  Scientific Collaboration}, \& {Virgo Collaboration}}]{abb2019}
{Abbott}, B.~P., {Abbott}, R., {Abbott}, T.~D., {et~al.} 2019{\natexlab{a}},
  \href{http://dx.doi.org/10.3847/2041-8213/ab3800}{\JournalTitle{\apjl}, 882,
  L24}

\bibitem[{{Abbott} {et~al.}(2019{\natexlab{b}}){Abbott}, {Abbott}, {Abbott},
  {Abraham}, {Acernese}, {Ackley}, {Adams}, {Adhikari}, {Adya}, {Affeldt},
  {Agathos}, {Agatsuma}, {Aggarwal}, {Aguiar}, {Aiello}, {Ain}, {Ajith},
  {Allen}, {Allocca}, {Aloy}, {Altin}, {Amato}, {Ananyeva}, {Anderson},
  {Anderson}, {LIGO Scientific Collaboration}, \& {Virgo
  Collaboration}}]{abb2019b}
---. 2019{\natexlab{b}},
  \href{http://dx.doi.org/10.1103/PhysRevX.9.031040}{\JournalTitle{PRX}, 9,
  031040}

\bibitem[{{Acernese} {et~al.}(2015){Acernese}, {Agathos}, {Agatsuma}, {Aisa},
  {Allemandou}, {Allocca}, {Amarni}, {Astone}, {Balestri}, {Ballardin},
  {Barone}, {Baronick}, {Barsuglia}, {Basti}, {Basti}, {Bauer}, {Bavigadda},
  {Bejger}, {Beker}, {Belczynski}, \& {Bersanetti}}]{acern2015}
{Acernese}, F., {Agathos}, M., {Agatsuma}, K., {et~al.} 2015,
  \href{http://dx.doi.org/10.1088/0264-9381/32/2/024001}{\JournalTitle{Classical
  and Quantum Gravity}, 32, 024001}

\bibitem[{{Ajith}(2011)}]{aji2011}
{Ajith}, P. 2011,
  \href{http://dx.doi.org/10.1103/PhysRevD.84.084037}{\JournalTitle{\prd}, 84,
  084037}

\bibitem[{{Antonini}(2013)}]{anto2013}
{Antonini}, F. 2013,
  \href{http://dx.doi.org/10.1088/0004-637X/763/1/62}{\JournalTitle{\apj}, 763,
  62}

\bibitem[{{Antonini} {et~al.}(2019){Antonini}, {Gieles}, \&
  {Gualandris}}]{anto2019}
{Antonini}, F., {Gieles}, M., \& {Gualandris}, A. 2019,
  \href{http://dx.doi.org/10.1093/mnras/stz1149}{\JournalTitle{\mnras}, 486,
  5008}

\bibitem[{{Antonini} \& {Perets}(2012)}]{antoper12}
{Antonini}, F., \& {Perets}, H.~B. 2012,
  \href{http://dx.doi.org/10.1088/0004-637X/757/1/27}{\JournalTitle{\apj}, 757,
  27}

\bibitem[{{Askar} {et~al.}(2017){Askar}, {Szkudlarek}, {Gondek-Rosi\'{n}ska},
  {Giersz}, \& {Bulik}}]{askar17}
{Askar}, A., {Szkudlarek}, M., {Gondek-Rosi\'{n}ska}, D., {Giersz}, M., \&
  {Bulik}, T. 2017,
  \href{http://dx.doi.org/10.1093/mnrasl/slw177}{\JournalTitle{\mnras}, 464,
  L36}

\bibitem[{{Banerjee}(2018)}]{baner18}
{Banerjee}, S. 2018,
  \href{http://dx.doi.org/10.1093/mnras/stx2347}{\JournalTitle{\mnras}, 473,
  909}

\bibitem[{{Banerjee} {et~al.}(2020){Banerjee}, {Belczynski}, {Fryer},
  {Berczik}, {Hurley}, {Spurzem}, \& {Wang}}]{banerjee2020}
{Banerjee}, S., {Belczynski}, K., {Fryer}, C.~L., {et~al.} 2020,
  \href{http://dx.doi.org/10.1051/0004-6361/201935332}{\JournalTitle{\aap},
  639, A41}

\bibitem[{{Bartos} {et~al.}(2017){Bartos}, {Kocsis}, {Haiman}, \&
  {M\'{a}rka}}]{bart17}
{Bartos}, I., {Kocsis}, B., {Haiman}, Z., \& {M\'{a}rka}, S. 2017,
  \href{http://dx.doi.org/10.3847/1538-4357/835/2/165}{\JournalTitle{\apj},
  835, 165}

\bibitem[{{Belczynski} {et~al.}(2010){Belczynski}, {Dominik}, {Bulik},
  {O'Shaughnessy}, {Fryer}, \& {Holz}}]{belcz2010}
{Belczynski}, K., {Dominik}, M., {Bulik}, T., {et~al.} 2010,
  \href{http://dx.doi.org/10.1088/2041-8205/715/2/L138}{\JournalTitle{\apjl},
  715, L138}

\bibitem[{{Belczynski} {et~al.}(2008){Belczynski}, {Kalogera}, {Rasio}, {Taam},
  {Zezas}, {Bulik}, {Maccarone}, \& {Ivanova}}]{belc2008}
{Belczynski}, K., {Kalogera}, V., {Rasio}, F.~A., {et~al.} 2008,
  \href{http://dx.doi.org/10.1086/521026}{\JournalTitle{\apjs}, 174, 223}

\bibitem[{{Belczynski} {et~al.}(2016){Belczynski}, {Repetto}, {Holz},
  {O'Shaughnessy}, {Bulik}, {Berti}, {Fryer}, \& {Dominik}}]{belcz2016}
{Belczynski}, K., {Repetto}, S., {Holz}, D.~E., {et~al.} 2016,
  \href{http://dx.doi.org/10.3847/0004-637X/819/2/108}{\JournalTitle{\apj},
  819, 108}

\bibitem[{{De Luca} {et~al.}(2020){De Luca}, {Desjacques}, {Franciolini},
  {Pani}, \& {Riotto}}]{deluca2020}
{De Luca}, V., {Desjacques}, V., {Franciolini}, G., {Pani}, P., \& {Riotto}, A.
  2020, \JournalTitle{arXiv e-prints}, arXiv:2009.01728

\bibitem[{{de Mink} \& {Mandel}(2016)}]{demi2016}
{de Mink}, S.~E., \& {Mandel}, I. 2016,
  \href{http://dx.doi.org/10.1093/mnras/stw1219}{\JournalTitle{\mnras}, 460,
  3545}

\bibitem[{{Dominik} {et~al.}(2015){Dominik}, {Berti}, {O'Shaughnessy},
  {Mandel}, {Belczynski}, {Fryer}, {Holz}, {Bulik}, \& {Pannarale}}]{domin2015}
{Dominik}, M., {Berti}, E., {O'Shaughnessy}, R., {et~al.} 2015,
  \href{http://dx.doi.org/10.1088/0004-637X/806/2/263}{\JournalTitle{\apj},
  806, 263}

\bibitem[{{Fishbach} \& {Holz}(2020)}]{fish2020}
{Fishbach}, M., \& {Holz}, D.~E. 2020, \JournalTitle{arXiv e-prints},
  arXiv:2009.05472

\bibitem[{{Fragione} \& {Bromberg}(2019)}]{frbr2019}
{Fragione}, G., \& {Bromberg}, O. 2019,
  \href{http://dx.doi.org/10.1093/mnras/stz2024}{\JournalTitle{\mnras}, 488,
  4370}

\bibitem[{{Fragione} {et~al.}(2018{\natexlab{a}}){Fragione}, {Ginsburg}, \&
  {Kocsis}}]{fgk2018}
{Fragione}, G., {Ginsburg}, I., \& {Kocsis}, B. 2018{\natexlab{a}},
  \href{http://dx.doi.org/10.3847/1538-4357/aab368}{\JournalTitle{\apj}, 856,
  92}

\bibitem[{{Fragione} {et~al.}(2018{\natexlab{b}}){Fragione}, {Ginsburg}, \&
  {Kocsis}}]{fraga2018}
---. 2018{\natexlab{b}},
  \href{http://dx.doi.org/10.3847/1538-4357/aab368}{\JournalTitle{\apj}, 856,
  92}

\bibitem[{{Fragione} {et~al.}(2019{\natexlab{a}}){Fragione}, {Grishin},
  {Leigh}, {Perets}, \& {Perna}}]{fragg2019}
{Fragione}, G., {Grishin}, E., {Leigh}, N. W.~C., {Perets}, H.~B., \& {Perna},
  R. 2019{\natexlab{a}},
  \href{http://dx.doi.org/10.1093/mnras/stz1651}{\JournalTitle{\mnras}, 488,
  47}

\bibitem[{{Fragione} \& {Kocsis}(2018)}]{fragk2018}
{Fragione}, G., \& {Kocsis}, B. 2018,
  \href{http://dx.doi.org/10.1103/PhysRevLett.121.161103}{\JournalTitle{\prl},
  121, 161103}

\bibitem[{{Fragione} \& {Kocsis}(2019)}]{fragk2019}
---. 2019,
  \href{http://dx.doi.org/10.1093/mnras/stz1175}{\JournalTitle{\mnras}, 486,
  4781}

\bibitem[{{Fragione} {et~al.}(2018{\natexlab{c}}){Fragione}, {Leigh},
  {Ginsburg}, \& {Kocsis}}]{fragb2018}
{Fragione}, G., {Leigh}, N. W.~C., {Ginsburg}, I., \& {Kocsis}, B.
  2018{\natexlab{c}},
  \href{http://dx.doi.org/10.3847/1538-4357/aae486}{\JournalTitle{\apj}, 867,
  119}

\bibitem[{{Fragione} {et~al.}(2019{\natexlab{b}}){Fragione}, {Leigh}, \&
  {Perna}}]{flp2019}
{Fragione}, G., {Leigh}, N. W.~C., \& {Perna}, R. 2019{\natexlab{b}},
  \href{http://dx.doi.org/10.1093/mnras/stz1803}{\JournalTitle{\mnras}, 488,
  2825}

\bibitem[{{Fragione} {et~al.}(2020){Fragione}, {Loeb}, \&
  {Rasio}}]{fragrasio2020}
{Fragione}, G., {Loeb}, A., \& {Rasio}, F.~A. 2020,
  \href{http://dx.doi.org/10.3847/2041-8213/ab9093}{\JournalTitle{\apjl}, 895,
  L15}

\bibitem[{{Fragione} \& {Silk}(2020)}]{frsilk2020}
{Fragione}, G., \& {Silk}, J. 2020, \JournalTitle{arXiv e-prints},
  arXiv:2006.01867

\bibitem[{{Fryer} {et~al.}(2012){Fryer}, {Belczynski}, {Wiktorowicz},
  {Dominik}, {Kalogera}, \& {Holz}}]{fryer2012}
{Fryer}, C.~L., {Belczynski}, K., {Wiktorowicz}, G., {et~al.} 2012,
  \href{http://dx.doi.org/10.1088/0004-637X/749/1/91}{\JournalTitle{\apj}, 749,
  91}

\bibitem[{{Fuller} \& {Ma}(2019)}]{fullerma2019}
{Fuller}, J., \& {Ma}, L. 2019,
  \href{http://dx.doi.org/10.3847/2041-8213/ab339b}{\JournalTitle{\apjl}, 881,
  L1}

\bibitem[{{Gayathri} {et~al.}(2020){Gayathri}, {Healy}, {Lange}, {O'Brien},
  {Szczepanczyk}, {Bartos}, {Campanelli}, {Klimenko}, {Lousto}, \&
  {O'Shaughnessy}}]{gaya2020}
{Gayathri}, V., {Healy}, J., {Lange}, J., {et~al.} 2020, \JournalTitle{arXiv
  e-prints}, arXiv:2009.05461

\bibitem[{{Gerosa} \& {Berti}(2019)}]{gerosa2019}
{Gerosa}, D., \& {Berti}, E. 2019,
  \href{http://dx.doi.org/10.1103/PhysRevD.100.041301}{\JournalTitle{\prd},
  100, 041301}

\bibitem[{{Giacobbo} \& {Mapelli}(2018)}]{giac2018}
{Giacobbo}, N., \& {Mapelli}, M. 2018,
  \href{http://dx.doi.org/10.1093/mnras/sty1999}{\JournalTitle{\mnras}, 480,
  2011}

\bibitem[{{Gonz{\'a}lez} {et~al.}(2007){Gonz{\'a}lez}, {Sperhake},
  {Br{\"u}gmann}, {Hannam}, \& {Husa}}]{gon07}
{Gonz{\'a}lez}, J.~A., {Sperhake}, U., {Br{\"u}gmann}, B., {Hannam}, M., \&
  {Husa}, S. 2007,
  \href{http://dx.doi.org/10.1103/PhysRevLett.98.091101}{\JournalTitle{Physical
  Review Letters}, 98, 091101}

\bibitem[{{Greene} {et~al.}(2019){Greene}, {Strader}, \& {Ho}}]{greene2019}
{Greene}, J.~E., {Strader}, J., \& {Ho}, L.~C. 2019, \JournalTitle{arXiv
  e-prints}, arXiv:1911.09678

\bibitem[{{G{\"u}ltekin} {et~al.}(2004){G{\"u}ltekin}, {Miller}, \&
  {Hamilton}}]{gultek2004}
{G{\"u}ltekin}, K., {Miller}, M.~C., \& {Hamilton}, D.~P. 2004,
  \href{http://dx.doi.org/10.1086/424809}{\JournalTitle{\apj}, 616, 221}

\bibitem[{{G{\"u}rkan} {et~al.}(2006){G{\"u}rkan}, {Fregeau}, \&
  {Rasio}}]{gurk2006}
{G{\"u}rkan}, M.~A., {Fregeau}, J.~M., \& {Rasio}, F.~A. 2006,
  \href{http://dx.doi.org/10.1086/503295}{\JournalTitle{\apjl}, 640, L39}

\bibitem[{{Hamers} \& {Samsing}(2019)}]{ham2019}
{Hamers}, A.~S., \& {Samsing}, J. 2019,
  \href{http://dx.doi.org/10.1093/mnras/stz1646}{\JournalTitle{\mnras}, 487,
  5630}

\bibitem[{{Harris}(1996)}]{harris1996}
{Harris}, W.~E. 1996,
  \href{http://dx.doi.org/10.1086/118116}{\JournalTitle{\aj}, 112, 1487}

\bibitem[{{Heger} {et~al.}(2003){Heger}, {Fryer}, {Woosley}, {Langer}, \&
  {Hartmann}}]{heger2003}
{Heger}, A., {Fryer}, C.~L., {Woosley}, S.~E., {Langer}, N., \& {Hartmann},
  D.~H. 2003, \href{http://dx.doi.org/10.1086/375341}{\JournalTitle{\apj}, 591,
  288}

\bibitem[{{Holley-Bockelmann} {et~al.}(2008){Holley-Bockelmann},
  {G{\"u}ltekin}, {Shoemaker}, \& {Yunes}}]{holl2008}
{Holley-Bockelmann}, K., {G{\"u}ltekin}, K., {Shoemaker}, D., \& {Yunes}, N.
  2008, \href{http://dx.doi.org/10.1086/591218}{\JournalTitle{\apj}, 686, 829}

\bibitem[{{Hurley} {et~al.}(2000){Hurley}, {Pols}, \& {Tout}}]{hurley2000}
{Hurley}, J.~R., {Pols}, O.~R., \& {Tout}, C.~A. 2000,
  \href{http://dx.doi.org/10.1046/j.1365-8711.2000.03426.x}{\JournalTitle{\mnras},
  315, 543}

\bibitem[{{Hurley} {et~al.}(2002){Hurley}, {Tout}, \& {Pols}}]{hurley2002}
{Hurley}, J.~R., {Tout}, C.~A., \& {Pols}, O.~R. 2002,
  \href{http://dx.doi.org/10.1046/j.1365-8711.2002.05038.x}{\JournalTitle{\mnras},
  329, 897}

\bibitem[{{Khan} {et~al.}(2016){Khan}, {Husa}, {Hannam}, {Ohme}, {P{\"u}rrer},
  {Forteza}, \& {Boh{\'e}}}]{khan2016}
{Khan}, S., {Husa}, S., {Hannam}, M., {et~al.} 2016,
  \href{http://dx.doi.org/10.1103/PhysRevD.93.044007}{\JournalTitle{\prd}, 93,
  044007}

\bibitem[{{Kremer} {et~al.}(2019){Kremer}, {Rodriguez}, {Amaro-Seoane},
  {Breivik}, {Chatterjee}, {Katz}, {Larson}, {Rasio}, {Samsing}, {Ye}, \&
  {Zevin}}]{krem2019}
{Kremer}, K., {Rodriguez}, C.~L., {Amaro-Seoane}, P., {et~al.} 2019,
  \href{http://dx.doi.org/10.1103/PhysRevD.99.063003}{\JournalTitle{\prd}, 99,
  063003}

\bibitem[{{Kremer} {et~al.}(2020){Kremer}, {Spera}, {Becker}, {Chatterjee}, {Di
  Carlo}, {Fragione}, {Rodriguez}, {Ye}, \& {Rasio}}]{krgap2020}
{Kremer}, K., {Spera}, M., {Becker}, D., {et~al.} 2020, \JournalTitle{arXiv
  e-prints}, arXiv:2006.10771

\bibitem[{{Liu} \& {Bromm}(2020)}]{liubro2020}
{Liu}, B., \& {Bromm}, V. 2020, \JournalTitle{arXiv e-prints}, arXiv:2009.11447

\bibitem[{{Liu} \& {Lai}(2018)}]{ll18}
{Liu}, B., \& {Lai}, D. 2018,
  \href{http://dx.doi.org/10.3847/1538-4357/aad09f}{\JournalTitle{\apj}, 863,
  68}

\bibitem[{{Liu} \& {Lai}(2020)}]{liulai2020}
---. 2020, \JournalTitle{arXiv e-prints}, arXiv:2009.10068

\bibitem[{{Lousto} {et~al.}(2010){Lousto}, {Campanelli}, {Zlochower}, \&
  {Nakano}}]{lou10}
{Lousto}, C.~O., {Campanelli}, M., {Zlochower}, Y., \& {Nakano}, H. 2010,
  \href{http://dx.doi.org/10.1088/0264-9381/27/11/114006}{\JournalTitle{Classical
  and Quantum Gravity}, 27, 114006}

\bibitem[{{Lousto} \& {Zlochower}(2008)}]{lou08}
{Lousto}, C.~O., \& {Zlochower}, Y. 2008,
  \href{http://dx.doi.org/10.1103/PhysRevD.77.044028}{\JournalTitle{\prd}, 77,
  044028}

\bibitem[{{Lousto} \& {Zlochower}(2011)}]{lou11}
---. 2011,
  \href{http://dx.doi.org/10.1103/PhysRevLett.107.231102}{\JournalTitle{Physical
  Review Letters}, 107, 231102}

\bibitem[{{Lousto} {et~al.}(2012){Lousto}, {Zlochower}, {Dotti}, \&
  {Volonteri}}]{lou12}
{Lousto}, C.~O., {Zlochower}, Y., {Dotti}, M., \& {Volonteri}, M. 2012,
  \href{http://dx.doi.org/10.1103/PhysRevD.85.084015}{\JournalTitle{\prd}, 85,
  084015}

\bibitem[{{Mapelli} {et~al.}(2020){Mapelli}, {Santoliquido}, {Bouffanais},
  {Arca Sedda}, {Giacobbo}, {Artale}, \& {Ballone}}]{mapell2020}
{Mapelli}, M., {Santoliquido}, F., {Bouffanais}, Y., {et~al.} 2020,
  \JournalTitle{arXiv e-prints}, arXiv:2007.15022

\bibitem[{{McKernan} {et~al.}(2020){McKernan}, {Ford}, \&
  {O'Shaughnessy}}]{mck2020}
{McKernan}, B., {Ford}, K.~E.~S., \& {O'Shaughnessy}, R. 2020,
  \JournalTitle{arXiv e-prints}, arXiv:2002.00046

\bibitem[{{O'Leary} {et~al.}(2016){O'Leary}, {Meiron}, \& {Kocsis}}]{olear2016}
{O'Leary}, R.~M., {Meiron}, Y., \& {Kocsis}, B. 2016,
  \href{http://dx.doi.org/10.3847/2041-8205/824/1/L12}{\JournalTitle{\apjl},
  824, L12}

\bibitem[{{O'Shaughnessy} {et~al.}(2010){O'Shaughnessy}, {Kalogera}, \&
  {Belczynski}}]{oshau2010}
{O'Shaughnessy}, R., {Kalogera}, V., \& {Belczynski}, K. 2010,
  \href{http://dx.doi.org/10.1088/0004-637X/716/1/615}{\JournalTitle{\apj},
  716, 615}

\bibitem[{{Pan} {et~al.}(2012){Pan}, {Loeb}, \& {Kasen}}]{pan2012}
{Pan}, T., {Loeb}, A., \& {Kasen}, D. 2012,
  \href{http://dx.doi.org/10.1111/j.1365-2966.2012.21030.x}{\JournalTitle{\mnras},
  423, 2203}

\bibitem[{{Planck Collaboration}(2016)}]{planck2016}
{Planck Collaboration}. 2016,
  \href{http://dx.doi.org/10.1051/0004-6361/201525830}{\JournalTitle{\aap},
  594, A13}

\bibitem[{{Portegies Zwart} {et~al.}(2004){Portegies Zwart}, {Baumgardt},
  {Hut}, {Makino}, \& {McMillan}}]{porte2004}
{Portegies Zwart}, S.~F., {Baumgardt}, H., {Hut}, P., {Makino}, J., \&
  {McMillan}, S. L.~W. 2004,
  \href{http://dx.doi.org/10.1038/nature02448}{\JournalTitle{\nat}, 428, 724}

\bibitem[{{Portegies Zwart} {et~al.}(2010){Portegies Zwart}, {McMillan}, \&
  {Gieles}}]{portegies2010}
{Portegies Zwart}, S.~F., {McMillan}, S. L.~W., \& {Gieles}, M. 2010,
  \href{http://dx.doi.org/10.1146/annurev-astro-081309-130834}{\JournalTitle{\araa},
  48, 431}

\bibitem[{{Rasskazov} \& {Kocsis}(2019)}]{rasskoc2019}
{Rasskazov}, A., \& {Kocsis}, B. 2019,
  \href{http://dx.doi.org/10.3847/1538-4357/ab2c74}{\JournalTitle{\apj}, 881,
  20}

\bibitem[{{Renzo} {et~al.}(2020){Renzo}, {Farmer}, {Justham}, {G{\"o}tberg},
  {de Mink}, {Zapartas}, {Marchant}, \& {Smith}}]{renzo2020}
{Renzo}, M., {Farmer}, R., {Justham}, S., {et~al.} 2020,
  \href{http://dx.doi.org/10.1051/0004-6361/202037710}{\JournalTitle{\aap},
  640, A56}

\bibitem[{{Rezzolla} {et~al.}(2008){Rezzolla}, {Barausse}, {Dorband},
  {Pollney}, {Reisswig}, {Seiler}, \& {Husa}}]{rezzolla2008}
{Rezzolla}, L., {Barausse}, E., {Dorband}, E.~N., {et~al.} 2008,
  \href{http://dx.doi.org/10.1103/PhysRevD.78.044002}{\JournalTitle{\prd}, 78,
  044002}

\bibitem[{{Rice} \& {Zhang}(2020)}]{rice2020}
{Rice}, J.~R., \& {Zhang}, B. 2020, \JournalTitle{arXiv e-prints},
  arXiv:2009.11326

\bibitem[{{Rodriguez} {et~al.}(2018){Rodriguez}, {Amaro-Seoane}, {Chatterjee},
  \& {Rasio}}]{rod18}
{Rodriguez}, C.~L., {Amaro-Seoane}, P., {Chatterjee}, S., \& {Rasio}, F.~A.
  2018,
  \href{http://dx.doi.org/10.1103/PhysRevLett.120.151101}{\JournalTitle{PRL},
  120, 151101}

\bibitem[{{Rodriguez} {et~al.}(2019){Rodriguez}, {Zevin}, {Amaro-Seoane},
  {Chatterjee}, {Kremer}, {Rasio}, \& {Ye}}]{rodetal2019}
{Rodriguez}, C.~L., {Zevin}, M., {Amaro-Seoane}, P., {et~al.} 2019,
  \href{http://dx.doi.org/10.1103/PhysRevD.100.043027}{\JournalTitle{\prd},
  100, 043027}

\bibitem[{{Rodriguez} {et~al.}(2020){Rodriguez}, {Kremer}, {Grudi{\'c}},
  {Hafen}, {Chatterjee}, {Fragione}, {Lamberts}, {Martinez}, {Rasio},
  {Weatherford}, \& {Ye}}]{rodr2020}
{Rodriguez}, C.~L., {Kremer}, K., {Grudi{\'c}}, M.~Y., {et~al.} 2020,
  \href{http://dx.doi.org/10.3847/2041-8213/ab961d}{\JournalTitle{\apjl}, 896,
  L10}

\bibitem[{{Romero-Shaw} {et~al.}(2020){Romero-Shaw}, {Lasky}, {Thrane}, \&
  {Calderon Bustillo}}]{romero2020}
{Romero-Shaw}, I.~M., {Lasky}, P.~D., {Thrane}, E., \& {Calderon Bustillo}, J.
  2020, \JournalTitle{arXiv e-prints}, arXiv:2009.04771

\bibitem[{{Safarzadeh} \& {Haiman}(2020)}]{safa2020}
{Safarzadeh}, M., \& {Haiman}, Z. 2020, \JournalTitle{arXiv e-prints},
  arXiv:2009.09320

\bibitem[{{Sakstein} {et~al.}(2020){Sakstein}, {Croon}, {McDermott},
  {Straight}, \& {Baxter}}]{sakst2020}
{Sakstein}, J., {Croon}, D., {McDermott}, S.~D., {Straight}, M.~C., \&
  {Baxter}, E.~J. 2020, \JournalTitle{arXiv e-prints}, arXiv:2009.01213

\bibitem[{{Samsing} {et~al.}(2018){Samsing}, {Askar}, \& {Giersz}}]{sams18}
{Samsing}, J., {Askar}, A., \& {Giersz}, M. 2018,
  \href{http://dx.doi.org/10.3847/1538-4357/aaab52}{\JournalTitle{\apj}, 855,
  124}

\bibitem[{{Spera} \& {Mapelli}(2017)}]{spera2017}
{Spera}, M., \& {Mapelli}, M. 2017,
  \href{http://dx.doi.org/10.1093/mnras/stx1576}{\JournalTitle{\mnras}, 470,
  4739}

\bibitem[{{Stone} {et~al.}(2017){Stone}, {Metzger}, \& {Haiman}}]{sto17}
{Stone}, N.~C., {Metzger}, B.~D., \& {Haiman}, Z. 2017,
  \href{http://dx.doi.org/10.1093/mnras/stw2260}{\JournalTitle{mnras}, 464,
  946}

\bibitem[{{The LIGO Scientific Collaboration} \& {the Virgo
  Collaboration}(2016)}]{ligo2016}
{The LIGO Scientific Collaboration}, \& {the Virgo Collaboration}. 2016,
  \href{http://dx.doi.org/10.3847/2041-8205/833/1/L1}{\JournalTitle{\apjl},
  833, L1}

\bibitem[{{The LIGO Scientific Collaboration} \& {the Virgo
  Collaboration}(2018)}]{ligo2018}
---. 2018,
  \href{http://dx.doi.org/10.1007/s41114-018-0012-9}{\JournalTitle{Living
  Reviews in Relativity}, 21, 3}

\bibitem[{{The LIGO Scientific Collaboration} \& {the Virgo
  Collaboration}(2020{\natexlab{a}})}]{ligo2020}
---. 2020{\natexlab{a}}, \JournalTitle{arXiv e-prints}, arXiv:2004.08342

\bibitem[{{The LIGO Scientific Collaboration} \& {the Virgo
  Collaboration}(2020{\natexlab{b}})}]{ligo2020c}
---. 2020{\natexlab{b}},
  \href{http://dx.doi.org/10.3847/2041-8213/ab75f5}{\JournalTitle{\apjl}, 892,
  L3}

\bibitem[{{The LIGO Scientific Collaboration} \& {the Virgo
  Collaboration}(2020{\natexlab{c}})}]{ligo2020new1}
---. 2020{\natexlab{c}}, \JournalTitle{arXiv e-prints}, arXiv:2009.01075

\bibitem[{{The LIGO Scientific Collaboration} \& {the Virgo
  Collaboration}(2020{\natexlab{d}})}]{ligo2020b}
---. 2020{\natexlab{d}},
  \href{http://dx.doi.org/10.3847/2041-8213/ab960f}{\JournalTitle{\apjl}, 896,
  L44}

\bibitem[{{The LIGO Scientific Collaboration} \& {the Virgo
  Collaboration}(2020{\natexlab{e}})}]{ligo2020new2}
---. 2020{\natexlab{e}},
  \href{http://dx.doi.org/10.3847/2041-8213/aba493}{\JournalTitle{\apjl}, 900,
  L13}

\bibitem[{{Vink} {et~al.}(2001){Vink}, {de Koter}, \& {Lamers}}]{vink2001}
{Vink}, J.~S., {de Koter}, A., \& {Lamers}, H.~J.~G.~L.~M. 2001,
  \href{http://dx.doi.org/10.1051/0004-6361:20010127}{\JournalTitle{\aap}, 369,
  574}

\bibitem[{{Woosley}(2017)}]{woosley2017}
{Woosley}, S.~E. 2017,
  \href{http://dx.doi.org/10.3847/1538-4357/836/2/244}{\JournalTitle{\apj},
  836, 244}

\end{thebibliography}

\end{document}